\documentclass[aps,prl,twocolumn,showpacs,amsmath,amssymb,floatfix]{revtex4}
\usepackage{graphicx}
\usepackage{bm}
\usepackage{epsfig}
\usepackage{array}

\def\be{\begin{equation}}
\def\ee{\end{equation}}
\def\bea{\begin{eqnarray}}
\def\eea{\end{eqnarray}}

\def\ra{\rangle}

\begin{document}

\title{Quantum Mutual Information Capacity for High Dimensional Entangled States}

\author{P.\ Ben Dixon}
\affiliation{Department of Physics and Astronomy, University of Rochester, Rochester, New York 14627, USA}

\author{Gregory A.\ Howland}
\affiliation{Department of Physics and Astronomy, University of Rochester, Rochester, New York 14627, USA}

\author{James Schneeloch}
\affiliation{Department of Physics and Astronomy, University of Rochester, Rochester, New York 14627, USA}

\author{John C.\ Howell}
\affiliation{Department of Physics and Astronomy, University of Rochester, Rochester, New York 14627, USA}

\date{\today}

\begin{abstract}
High dimensional Hilbert spaces used for quantum communication channels offer the possibility of large data transmission capabilities. We propose a method of characterizing the channel capacity of an entangled photonic state in high dimensional position and momentum bases.  We use this method to measure the channel capacity of a parametric downconversion state, achieving a channel capacity over 7 bits/photon in either the position or momentum basis, by measuring in up to 576 dimensions per detector.  The channel violated an entropic separability bound, suggesting the performance cannot be replicated classically.
\end{abstract}

%
\pacs{42.50.Dv, 03.67.Bg, 03.67.Hk}
\maketitle

\section{Introduction}
Quantum systems can be entangled in various degrees of freedom.  Typical examples include photonic polarization states \cite{PolEnt} and atomic or ionic energy levels \cite{IonEnt}.  These systems enable various technologies such as quantum communication, quantum cryptography, and quantum computation \cite{QCompComm, QCrypt, QComp}, however the systems are not in principle limited to two states \cite{OAMDim, OAMHighDim, TimeEnergyHighDim, FractionalFourierHighDim, SpatialQudits1, SpatialQudits2}.  Indeed, for communication purposes---such as quantum key distribution \cite{QKD} or dense coding communication \cite{QComm}---higher dimensional states increase quantum communication channel capacity and offer additional benefits such as increased security \cite{LargeDimBell, LargeDimSec}.  The photonic position degree of freedom is a good candidate for practical high-dimensional entangled systems due to the wide availability of off the shelf technology for manipulating this degree of freedom \cite{SpatialEase}.  We propose a quantum channel capacity characterization that considers both the quantum state and the measurement apparatus.  We use this method to measure the channel capacity of a high dimensional position and momentum entangled photonic state.  Our measurements include up to 576 dimensions per detector and demonstrate channel capacities of over 7 bits/photon.

The capacity of a quantum channel using entangled photons characterizes information transfer in joint detection events.  The locations where joint photon detections can take place can be thought of as characters in an alphabet; the size of this alphabet is the number of distinguishable joint detection locations available within the beam envelopes (see for example \cite{ShannonTheory}).  Theoretically, the channel capacity calculation the best possible measurement for a given state used as the channel. Practically however, the measurement technique is not necessarily optimal and the channel capacity depends on the details of this measurement.  


Experimental characterization of channel capacities has consisted of performing quantum state tomography and then using the result in channel capacity calculations.  Recently Pors {\it et al}. \cite{ShannonDim} proposed a more direct way of characterizing the channel capacity without recreating the full quantum state.  By considering the quantum state in conjunction with the measurement apparatus, they define an effective channel dimensionality called the Shannon dimensionality.  We propose a similar measure: an effective entropic channel capacity---rather than dimensionality---that considers both the state and the measurement apparatus, and measures bits of information per detection event.  This quantity characterizes the information capacity that can be effectively probed for a given system.  It depends solely on each party's measurements and is independent of character coding scheme.


\section{Theoretical Description}
\begin{figure}
\includegraphics[scale=0.25]{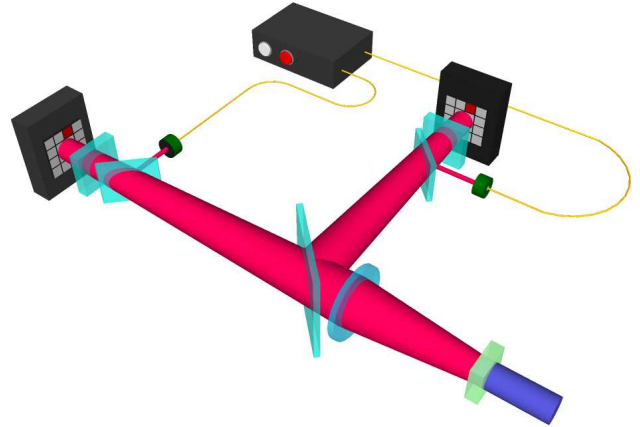}
\caption{Experimental Setup.  A collimated laser beam undergoes spontaneous parametric down-conversion at a nonlinear crystal.  The output passes a focusing lens followed by a beam-splitter.  The outputs from the beam-splitter are sent to digital micro-mirror devices at either image planes or Fourier planes of the crystal.  The micro-mirror devices are set to retro-reflect the beams, a quarter wave plate and a polarizing beam-splitter send the retro-reflected beam to a single-photon detector.  A coincidence circuit correlates these measurements. }
\label{Experiment}
\end{figure}
We use mutual information to quantify the channel capacity.  Mutual information describes how much information can be determined about a random variable \(A\), by knowing the value of a correlated random variable \(B\) \cite{ShannonTheory, OneTimePad}.  Variables \(A\) and \(B\) are characterized by the values they take \(a\) and \(b\), respectively, and the probability of these values \(p(a)\) and \(p(b)\), respectively.  The mutual information can be written as
\be
I(A;B) = H(A) + H(B) - H(A,B),
\label{MutInfo}
\ee
where, for example
\be
H(A) = -\sum_{a \in A} p(a) \log p(a)
\label{MargEnt}
\ee
is the marginal entropy of \(A\), and
\be
H(A,B) = -\sum_{\substack{a \in A \\ b \in B}} p(a,b) \log p(a,b)
\label{JointEnt}
\ee
is the joint entropy of \(A\) and \(B\).  The function \(p(a,b)\) is the joint probability distribution which characterizes the correlation between \(A\) and \(B\).

We created a position-momentum entangled state using spontaneous parametric-downconversion (SPDC). The state, represented both in position and in momentum, is approximated as \cite{state}
\be
\begin{split}
|\psi\ra &= \int \mathrm{d}\vec{x}_a \mathrm{d}\vec{x}_b \,f(\vec{x}_a, \vec{x}_b) \,\hat{a}^{\dagger}_a \hat{a}^{\dagger}_b \,|0\ra\\
         &= \int \mathrm{d}\vec{k}_a \mathrm{d}\vec{k}_b \,\tilde{f}(\vec{k}_a, \vec{k}_b) \,\hat{a}^{\dagger}_a \hat{a}^{\dagger}_b \,|0\ra
\label{StatePosMom}
\end{split}
\ee
where \(\hat{a}^{\dagger} \) is the photon creation operator.  Subscript \(a\) or \(b\) indicates the photon is created in the signal or idler mode, which are sent to Alice or Bob, respectively.  The function
\be
f(\vec{x}_a, \vec{x}_b) = N \exp \left( \frac{-(\vec{x}_a-\vec{x}_b)^2}{4 \sigma_c^2}\right) \exp \left(\frac{-(\vec{x}_a+\vec{x}_b)^2}{16 \sigma_p^2} \right)
\label{StateFuncPos}
\ee
is the entangled biphoton wavefunction in the position basis, and
\be
\begin{split}
\tilde{f}(\vec{k}_a, \vec{k}_b) = (4 \sigma_p \sigma_c)^2 & N  \exp ( - \sigma_c^2 (\vec{k}_a-\vec{k}_b)^2)\\
 \times &\exp ( - 4 \sigma_p^2 (\vec{k}_a+\vec{k}_b)^2 )
\label{StateFuncMom}
\end{split}
\ee
is the biphoton wavefunction in the momentum basis.  In these equations \(N = (2 \pi \sigma_p \sigma_c)^{-1} \) is a normalization constant.  These representations are related through a Fourier transform, which inverts the relative Gaussian function widths, such that position correlations become momentum anti-correlations.

To measure position correlations we put spatially-resolving single-photon detectors at image planes of the SPDC source: to measure momentum correlations we put the detectors at Fourier transform planes of the source.  For our purposes then, random variable \(A\) corresponds either to the position or momentum of Alice's photon and \(B\) corresponds either to the position or momentum of Bob's photon.

The theoretical maximum mutual information for the wavefunction in Eq.\ \ref{StateFuncPos} (measuring in the position basis), is:
\be
I(A;B) =
-\int p(\vec{x}_a,\vec{x}_b) \log \left(\frac{p(\vec{x}_a,\vec{x}_b)}{p(\vec{x}_a)p(\vec{x}_b)} \right) \mathrm{d} \vec{x}_a \mathrm{d} \vec{x}_b
\label{MutInfoContFormula}
\ee
where \(p(\vec{x}_a,\vec{x}_i) = |f(\vec{x}_a,\vec{x}_b)|^2 \) and \( p(\vec{x}_a) = \int |f(\vec{x}_a,\vec{x}_b)|^2 \mathrm{d}\,\vec{x}_b. \)
For the momentum basis, the same relations hold, but the position variables are replaced by the momentum variables and the position wavefunction is replaced by the momentum wavefunction of Eq.\ \ref{StateFuncMom}.

For either basis, this theoretical maximum simplifies to
\be
I(A;B) = \log\left( \frac{4 \sigma_p^2 + \sigma_c^2}{4 \, \sigma_c\sigma_p} \right)^2,
\label{MutInfoCont}
\ee
which is independent of detector characteristics.  In the limit of strong correlations \( (\sigma_p / \sigma_c) \gg 1 \) the mutual information reduces to \( I(A;B) = \log\left(\sigma_p / \sigma_c \right)^2 \).  The ratio \(\sigma_p / \sigma_c\) is the familiar Fedorov ratio for quantifying entanglement \cite{Fedorov}.  Our physical SPDC state had a beam envelope width of \(\sigma_p = 1500\) \(\mu\)m and a correlation width of approximately \( \sigma_c = 40\) \(\mu\)m, resulting in an optimum mutual information from Eq.\ \ref{MutInfoCont} of \(I \cong 10\) bits/photon.

The measurement apparatus consisted of a digital micromirror device (DMD) chip reflecting a portion of the signal or idler beam onto a single photon counting module.  The DMD chip allowed us to raster scan over the face of the beam in a controllable number of detection pixels, giving varying detector resolution.  To incorporate the effects of the measurement apparatus, we integrate the probability density over the pixel area.  For position correlations between the \(m^{th}\) pixel on Alice's detector, and the \(n^{th}\) pixel on Bob's detector, the joint detection probability is:
\be
p(m,n)=\int_{m} \mathrm{d}\vec{x}_a \int_{n} \mathrm{d}\vec{x}_b \,|f(\vec{x}_a,\vec{x}_b)|^2.
\ee

Similarly, for momentum correlations between the \(m^{th}\) pixel on Alice's detector, and the \(n^{th}\) pixel on Bob's detector, the joint detection probability is:
\be
p(m,n)=\int_{m} \mathrm{d}\vec{k}_a \int_{n} \mathrm{d}\vec{k}_b \,|\tilde{f}(\vec{k}_a,\vec{k}_b)|^2.
\ee

The detected mutual information in either position or momentum is then
\be
\begin{split}
I(A;B) =&
\sum_{m} p(m) \log p(m)
+\sum_{n} p(n) \log p(n)\\
&-\sum_{m,n} p(m,n) \log p(m,n) ,
\label{MutInfoPixelFormula}
\end{split}
\ee
where, for example,
\be
p(m) = \sum_{n}p(m,n)
\ee
is the marginal probability for a pixel on Alice's detector.

\begin{figure*}
\includegraphics[scale=.87]{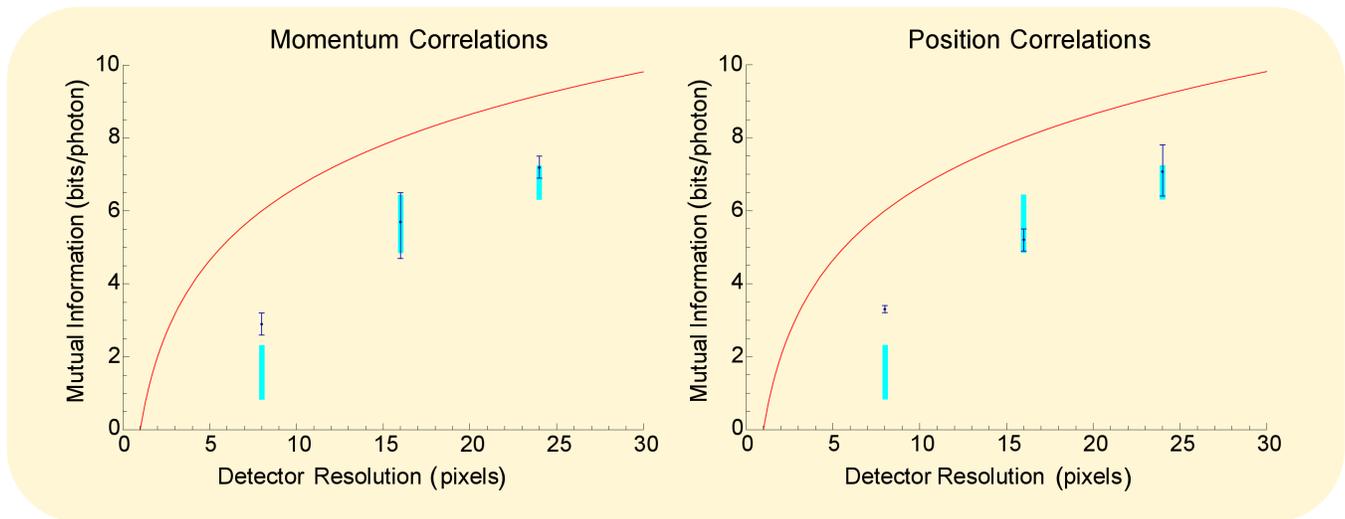}
\caption{Mutual Information Data.  Mutual information for position correlation measurements and momentum correlation measurements are shown as a function of detector resolution.  Data for detector resolutions of \(8\times8\) pixels, \(16\times16\) pixels, and \(24\times24\) pixels are shown.  The dark blue points with error bars are experimental data.  The light blue bars are numerical simulations based on Eq.\ \ref{MutInfoPixelFormula} both for the case of perfectly relative transversely aligned detectors and the case of a relative transverse misalignment of half a pixel.  The red curve is the maximum mutual information that can be detected for the number pixels per detector. }
\label{MutInfoPlot}
\end{figure*}
It should be noted that the mutual information calculation only takes into account coincident photon detections.  Thus a non-ideal detection efficiency does not change the form of the mutual information equations, rather it reduces the effective SPDC beam intensity.  This effect reduces the system's information per unit time, but not the information per detected photon pair.

\section{Experiment}
The experimental setup is shown in Fig.\ \ref{Experiment}.  A 325 nm wavelength laser beam with diameter a diameter of \(\sigma_p = 1500\) \(\mu\)m pumped a 10 mm long BBO nonlinear crystal aligned for Type-I degenerate collinear SPDC. The SPDC output had a correlation width of approximately \( \sigma_c = 40\) \(\mu\)m.  For these parameters the optimum mutual information from Eq.\ \ref{MutInfoContFormula} is \(I \cong 10\) bits/photon.  For measuring position correlations the SPDC output passed through a 125 mm focal length lens followed by a beamsplitter and a spatially resolving detectors were located at the resulting image planes of the crystal.  For measuring momentum correlations the SPDC output passed through a 150 mm focal length lens followed by a beamsplitter and a spatially resolving detector was placed at the resulting Fourier transform planes of the crystal.

The spatially resolving single-photon detectors consisted of a computer controlled digital micro-mirror device from Texas Instruments in conjunction with a Perkin Elmer single photon avalanche diode (SPAD) running in geiger mode.  The micro-mirror displays had \(1064 \times 768\) resolution which selectively reflected portions of the SPDC signal and idler beams to SPADs.  Photon detection events were correlated with a PicoHarp 300 from PicoQuant with a 3 ns coincident window.

The micro-mirror displays were each raster scanned and counts for each pixel pair were recorded for between 1 and 5 seconds.  These double raster scans were set such that they were centered on the signal and idler beams, divided into \(8 \times 8 \) pixels, \(16 \times 16 \) pixels, and \(24 \times 24 \) pixels, encompassing \(~80\%\) of each beam.

For ideal alignment, a given pixel on Alice's detector will correlate very well to only one pixel on Bob's detector.  In practice however, a relative lateral shift of pixels between Alice's and Bob's detectors---both vertically and horizontally---spreads correlations to four pixels at best.  However, pixels far from the correlated pixel will still have no correlation.  This was verified experimentally and it allowed us, for a given pixel on Alice's detector, to scan only in a region of interest around the correlated pixel on Bob's detector, thus reducing the time required to complete a double raster scan.

\section{Results}
Both the predicted and experimentally measured values for mutual information are shown in Fig.\ \ref{MutInfoPlot}.  Mutual information values for both position correlation measurements and momentum correlation measurements are presented.  Uncertainties in detected photon number \(N\) for each point in the double raster scan were assumed to be \( \sqrt N \).  This uncertainty was then propagated through the entropy calculations, giving the uncertainties for the measured mutual information values. These values were found to be in agreement with the statistics found by taking multiple data scans for a given detector resolution.  It should be noted that this uncertainty calculation method does not take into account detector dark counts. Since the dark counts from each detector are uncorrelated, the dark coincident rate is much less than the coincident rate from the highly correlated SPDC state.

Light blue bars represent predicted mutual information values from numerical calculations of Eq.\ \ref{MutInfoPixelFormula}.  The tops of the bars correspond to perfect lateral pixel alignment between Alice and Bob, and the bottoms correspond to relative lateral shifts, both horizontally and vertically, of half a pixel.  These cases represent the maximum and minimum mutual information possible for a given number of detector pixels.  The dark blue circles represent experimentally measured channel capacities.  The red curve gives the maximum mutual information that can be detected \(I = \log(N) \) for \(N\) pixels per detector.  Data for the detectors scanning in \(8 \times 8\) pixels, \(16 \times 16\) pixels, and \(24\times24\) pixels are shown.  The experimental data agrees well with the theoretically predicted values.

For momentum correlations, a maximum mutual information of \(7.2 \pm 0.3\) bits/photon was achieved; for position correlations only \(7.1 \pm 0.7\) bits/photon were achieved.  In principle, the two measurement bases should give the same mutual information.  However, the alignment for position correlations was more sensitive---the reduction of mutual information for this basis most likely resulted from slight system misalignments.

The 576 dimensional measurement space is 16 times larger than the previous maximum for position-momentum entangled photons, and had an increase of bit capacity of more than \(50\%\) \cite{SpatialQudits2}.  It should be noted that channel capacity characterization is different from using the channel for communication. When used for key distribution or communication, the characterized channel will indeed transmit 7 bits of information for a single joint detection event, despite the fact that our characterization method requires many photon detection events.  The use of this channel for key distribution or communication does however require some additional structure \cite{QKD, QComm}.  We are further investigating the ultimate experimental realization these structures.  Although our entropic channel characterization measure is similar to the Shannon dimensionality of Pors {\it et al}.\, it has several important differences: the units of the measures are not the same and the the weighting of the different pixel probabilities differs between the measures---however for a flat probability distribution the measures have identical magnitudes.  By measuring bits of information rather than dimensionality, our measure is more directly linked to channel information capacity.

\section{Nonclassicality}
By taking data in two mutually unbiased bases (position and momentum) we are able to test the separability of the state used as the communication channel.  A separable state satisfies the inequality \( H(A|B)_P + H(A|B)_M \geq \log_2( \pi e) \approx 3.09\) where subscripts \(P\) or \(M\) indicate measurements in the position or momentum bases respectively, and \( H(A|B) = H(A,B) - H(B) \) is the conditional entropy of \(A\) given \(B\) \cite{EntropicSep1,EntropicSep2}.  From the \(8 \times 8\) pixel scan data, we calculate
\begin{eqnarray}
H(A|B)_P + H(A|B)_M &=& 3.7 \pm 0.3\\
H(B|A)_P + H(B|A)_M &=& 3.9 \pm 0.2,
\end{eqnarray}
from the \(16 \times 16\) pixel scan data, we calculate
\begin{eqnarray}
H(A|B)_P + H(A|B)_M &=& 3.2 \pm 0.8\\
H(B|A)_P + H(B|A)_M &=& 3.1 \pm 0.8,
\end{eqnarray}
and from the \(24 \times 24\) pixel scan data, we calculate
\begin{eqnarray}
H(A|B)_P + H(A|B)_M &=& 2.2 \pm 0.7\\
H(B|A)_P + H(B|A)_M &=& 2.2 \pm 0.6.
\end{eqnarray}
As the scan resolution increases, we probe stronger correlations and the separability sums decrease.  For \(16\times16\) resolution the separability bound is approached.  For \(24\times24\) resolution the separability bound is violated by more than 1.3 standard deviations, indicating it is unlikely that the channel performance can be replicated classically.

Although our characterization method is independent of character coding scheme, it suggests a simple one:  we assign an alphabet character to each of the pixels.  This scheme achieves the measured channel capacities, and errors can be minimized by reducing the size (but not location) of pixels on Alice's detector, such that the system is unaffected by small relative lateral pixel misalignments between Alice and Bob.

Our experiment was limited only by pump laser flux.  Higher resolution scans are in principle possible, however they are not practical for our setup due to the scan time exceeding the relaxation time of the optical setup.  A more powerful pump laser would enable higher resolution scans while maintaining feasible times.  Such scans would come closer to experimentally demonstrating the optimum capacity for a position momentum entangled state.

\section{Concluding Remarks}
We have proposed and demonstrated a simple method of characterizing the quantum mutual information based channel capacity of a quantum communication channel using position and momentum entangled photons and a controllable pixel mirror.  We measured up to 576 dimensions per detector, in both the position and the momentum basis, which resulted in a measured channel capacity of more than 7 bits/photon for either basis.  The channel violated an entropic separability bound, suggesting the performance cannot be replicated classically.

We acknowledge discussions with B.\ I.\ Erkmen and support from DARPA DSO InPho grant W911NF-10-1-0404 and USARO MURI grant W911NF-05-1-0197.

\bibliography{MutInfo}

\end{document}